# An electrical probe of the phonon mean-free path spectrum

Ashok T. Ramu[1], Nicole I. Halaszynski[1], Jonathan D. Peters[1], Carl D. Meinhart[2] and John E. Bowers[1]

1. Department of Electrical and Computer Engineering, University of California Santa Barbara, Santa Barbara, California, CA-93106, USA.
2. Department of Mechanical Engineering, University of California Santa Barbara, Santa Barbara, California, CA-93106, USA.

*Most studies of the mean-free path accumulation function (MFPAF) rely on optical techniques to probe heat transfer at length scales on the order of the phonon mean-free path. In this paper, we propose and implement a purely electrical probe of the MFPAF that relies on photo-lithographically defined heater-thermometer separation to set the length scale. An important advantage of the proposed technique is its insensitivity to the thermal interfacial impedance and its compatibility with a large array of temperature-controlled chambers that lack optical ports. Detailed analysis of the experimental data based on the enhanced Fourier law (EFL) demonstrates that heat-carrying phonons in gallium arsenide have a much wider mean-free path spectrum than originally thought.*

1. Introduction

The mean-free path accumulation function (MFPAF) introduced by Dames and Chen [1] is a powerful tool in studying ballistic phonon transport. The MFPAF at a given mean-free path Λ is defined as the effective thermal conductivity (ETC) contributed by all phonons with mean-free paths less than or equal to Λ. The utility of the MFPAF lies in that it explains within a unified framework[2] diverse experiments, like the transient gratings[3], time-domain thermoreflectance (TDTR)[4][5] and frequency domain thermoreflectance (FDTR)[6] experiments, that probe heat transport on length scales comparable to the phonon mean-free path.

For bulk materials, measurements of the MFPAF have been conducted for crystalline[7] and amorphous[8] materials. For nanostructured materials, Yang and Dames [9] have given a relationship to the MFPAF of bulk materials. The MFPAF of nanostructured materials has been determined by measuring length-dependent conductivity in nanowires[10]. The MFPAF has been applied to the determination of thermal properties of nanostructured materials from bulk properties [11]. The concept of MFPAF has been applied to the study of thermal interfaces as well [12].

We propose and implement a novel experiment to deduce the MFPAF from purely electrical measurements. We propose to excite a metal line deposited on the material to be characterized with an alternating current with frequency in the range of 200-2000 Hz – please see Fig. 1. The AC voltage developed on the heater is measured across the inner pads of the heater in order to deduce the power input given the line resistance. A temperature sensing line is deposited a short distance from the heater line. The distance between the heater and thermometer would be on the order of the mean-free path of dominant low-frequency modes (500 nm – 4000 nm for bulk Si and GaAs) [13]

The temperature of the thermometer line is extracted by passing a direct current through its outer pads and measuring the second harmonic voltage across its inner pads (Fig. 1). This is a variant of the AC third-harmonic technique originally invented by Atalla and coworkers [14] and popularized by

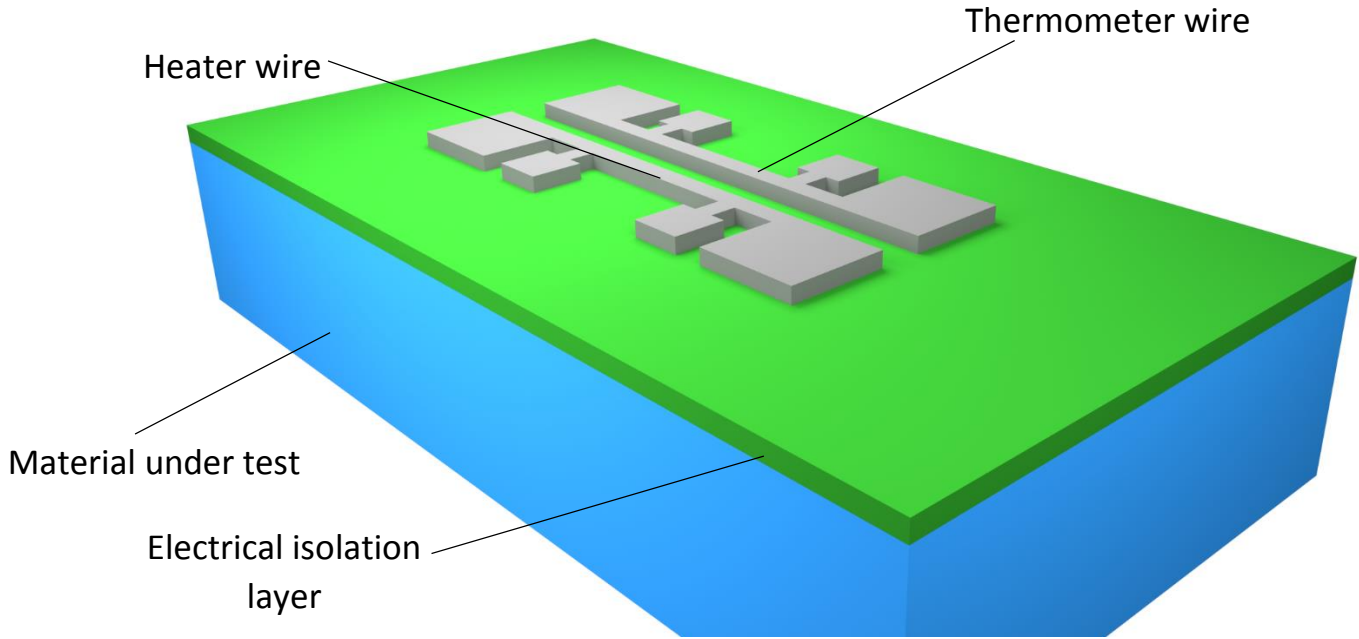

Fig. 1: Schematic of proposed experiment to investigate non-Fourier heat transfer

Cahill and coworkers [15] under the label "3-omega method". We note that our "2-omega" method was originally developed to measure the thermal anisotropy of bulk materials [16-17].

The overall experimental procedure is as follows: First we conduct the standard 3-omega experiment using one of the two identical lines, heater or thermometer, and derive the bulk thermal conductivity using the slope method. The details of this experiment are described in [15] and we shall not repeat it here. Next, we excite the heater with an AC current and pass a DC current through the thermometer. If the AC current has a frequency $f$, the Joule power has a frequency $2f$, which creates temperature oscillations (TOs) at a frequency $2f$. The TO results in a thermometer resistance oscillation at a frequency $2f$ because of the temperature coefficient of resistance of the thermometer metal line. This resistance oscillation is converted by the DC thermometer current into a voltage oscillation at a frequency $2f$ which is measured using a lock-in amplifier.

We scale the heater input power to 1 W per cm heater length, and determine the difference between the data and Fourier law simulation (using the measured value of the bulk thermal conductivity) at 2000 Hz. This frequency is used because it is high enough to minimize finite-length effects (thus enabling a 2D analysis) and to minimize substrate effects, yet low enough for the signal strength to be adequate (please see Sec. 4, especially Fig. 6, for justification of this choice of frequency). We repeat this procedure for heater-thermometer pairs of varying widths, and plot the deviation from the Fourier law vs. the heater-thermometer separation. We fit this graph to a two-channel model [18] and deduce the model parameters, and thence the MFPAF of the material being investigated. Subsequent sections provide details of the experimental procedure, and describe the results obtained for GaAs and strontium titanate.

2. Theoretical framework

We analyze the experimental data in the context of the two-channel enhanced Fourier law (EFL) developed earlier [18][19]. We desire the temperature profile on a thermometer of width $w_t$ separated from a heater of width $w_h$ by a center-to-center distance of *D*. All variables (excluding parameters) are assumed to have an implicit time dependence $e^{-j2\omega t}$. First we state the enhanced Fourier law (EFL) in three dimensions as derived in [18], ignoring circulation of the heat-flux:

$$\boldsymbol{q}^{HF} = -\kappa^{HF}\nabla T \tag{1}$$

$$\boldsymbol{q}^{LF} = -\kappa^{LF}\nabla T + \frac{3}{5}\Lambda^2 \nabla(\nabla \cdot \boldsymbol{q}^{LF}) \tag{2}$$

$\boldsymbol{q}^{HF}$ is the high-frequency (HF) heat-flux that follows the Fourier law, $\kappa^{HF}$ is the corresponding thermal conductivity, $\boldsymbol{q}^{LF}$ is the heat-flux in the low-frequency (LF) quasi-ballistic mode, $\Lambda$ is the mean-free path of the LF mode, $\kappa^{LF}$ is the kinetic theory value of the modal thermal conductivity, and *T* is the temperature. We note that $\kappa^{HF} + \kappa^{LF} = \kappa^{bulk}$ (the bulk thermal conductivity) so that the only independent parameters in this model are $\kappa^{LF}$ and $\Lambda$.

Next we state the law of energy conservation:

$$\nabla \cdot (\boldsymbol{q}^{HF} + \boldsymbol{q}^{LF}) = -j2\omega C_v T \tag{3}$$

$\omega$ is the angular frequency of the heater current and $C_v$ is the heat capacity of all phonon modes (HF and LF).

The boundary conditions are given by,

$$\boldsymbol{q}^{HF} \cdot \hat{\boldsymbol{n}} = -P \tag{4}$$

$$\boldsymbol{q}^{LF} \cdot \hat{\boldsymbol{n}} = 0 \tag{5}$$

Here *P* is the power fluxed by the heater per unit area, calculated using a total power assumed henceforth to be fixed at 1 W per cm heater length. All raw experimental data is scaled to this input power before presenting in this paper.

Equations (1-5) are solved numerically for the geometry of Fig. 1 using the finite-element method implementation COMSOL®. The deviation of the two-channel EFL simulation result from the Fourier law is noted at each experimental heater-thermometer separation *D*, and parameters $\kappa^{LF}$ and $\Lambda$ are varied iteratively until the simulated deviation closely matches the experimental deviational data in the least mean-square sense. COMSOL®'s optimization solver is used for this purpose.

Once $\kappa^{LF}$ and $\Lambda$ are known, an approximation for the mean-free path accumulation function may be obtained as shown in Fig. 4. We neglect the thin electrical isolation layer as well as the finite thickness of the heater in our analysis of the '2-omega' experiment; we consider the effect of these approximations in Sec. 4.

3. Experimental methodology

On the GaAs substrate, 50 nm silicon dioxide is deposited using plasma-enhanced chemical vapor deposition. Aluminium heater-thermometer pairs with each line of length 300 microns, width varying as 500nm, 1000 nm, and 2000 nm, and thickness of 1.2 microns, are fabricated on top of the oxide on GaAs, with a 23 nm titanium film acting as an adhesion layer. Thus the oxide layer electrically isolates the heater/thermometer pair from the unintentionally conducting GaAs substrate. An array of devices is made with heater-thermometer separations ranging over 1000, 1200, 1500 and 2500 nm. After initial 3-omega experiments to yield the bulk thermal conductivity, the heater/thermometer widths are fixed at 500 nm (nominally) for the 2-omega experiment, and their separation *D* alone is varied over the range mentioned above.

On strontium titanate (STO) substrate, the process is slightly different. The 50 nm oxide layer is deposited using radio-frequency sputtering, and the metallization consists of Ti/Au/Ti/Au 20/10/20/~500 nm. The rest of the dimensions are nominally the same. Process details are provided in the appendix.

The dimensions mentioned above are nominal, as laid out on the photolithography mask. Actual dimensions were measured using scanning electron microscopy (SEM) on each device that was probed, and the measured values were used in analyses. The width of the nominally 500 nm wide lines was actually 700 nm on an average, with deviations of +/- 100 nm depending on the device. The sensitivity of the modeling to the heater/thermometer widths is very small; for example varying the heater and thermometer widths from 650 nm to 800 nm resulted in only a <5% change in the Fourier law result for the temperature of the thermometer on GaAs, for a constant heater-thermometer center-to-center separation of 1096 nm. Thus we can plot data points collected on devices with different heater and thermometer widths on the same graph. The mean measured center-center heater-thermometer separations were 1096, 1292, 1566 and 2609 nm. These separations showed much less spread from device to device (only +/- 10 nm) than the individual line widths.

For implementation of the "2-omega" method, a 5 V-rms sinusoidal waveform generator feeds the outer pads of the heater through a 100 Ohm resistor. A 10 V DC source feeds the outer pads of the thermometer through a 1 kilo-Ohm resistor. The resistors are included to reduce the noise introduced into the circuit by the sources. The nominally 500 nm wide lines have a resistance of 20-40 Ohm. The voltage across the inner pads of the heater is noted by a lock-in amplifier, in order both to monitor the power input as well as to provide a reference phase for detection. The second harmonic of the voltage across the inner pads of the thermometer is noted by the lock-in amplifier, and converted into a temperature oscillation amplitude. The component of the thermometer second harmonic voltage in phase with the heater power alone is used, since the out-of-phase signal is much weaker. Some typical values for various voltages are: 0.8 V-rms for the AC voltage across the inner pads of the heater, 0.3 V for the DC voltage across the inner pads of the thermometer, and 1.2 millivolt-rms and 200 microvolts-rms for the second harmonic voltage across the inner pads of the thermometer in phase and out of phase with the heater power respectively, at a frequency of 2000 Hz. The second harmonic voltage across the inner pads of the thermometer is converted into a temperature oscillation amplitude using the relation [17]:

$$T_{meas}^{phase} = \frac{\sqrt{2}L_{cm}}{\alpha I_{dc}} \cdot \frac{V_{2\omega}^{phase}}{V_{1\omega}^2} \qquad (6)$$

Here $L_{cm}$ is the length of the heater in cm, $\alpha$ is the temperature coefficient of resistance of the thermometer, $I_{dc}$ is the DC current through the thermometer, $V_{2\omega}^{phase}$ is the second harmonic voltage across the inner pads of the thermometer in phase with the heater power, and $V_{1\omega}$ is the AC voltage across the inner pads of the heater line.

Three to five devices (i.e. heater-thermometer pairs) were measured for each value of the separation in order to establish the repeatability of the experimental technique, and the results presented in the next section were fit to the two-channel enhanced Fourier law (EFL) implemented in COMSOL®, using COMSOL®'s optimization solver. Input parameters to the model are summarized in Table 1.

Table 1: Material parameters used in the modeling of GaAs and STO substrates

| Parameter | Value | Source |
| --- | --- | --- |
| $\kappa^{bulk}$ for GaAs | 60 W/m-K | '3-omega' measurement |
| $\kappa^{bulk}$ for STO | 12.5 W/m-K | '3-omega' measurement |
| Heat capacity $C_v$, GaAs | $1.76 \times 10^6$ J/m$^3$-K | [22] |
| Heat capacity $C_v$, STO | $2.75 \times 10^6$ J/m$^3$-K | [23] |
| $\alpha$ for Ti/Al 23/1200 nm | $3.7 \times 10^{-3}$ K$^{-1}$ | I-V-T measurements |
| $\alpha$ for Ti/Au/Ti/Au 20/10/20/500 nm | $3.3 \times 10^{-3}$ K$^{-1}$ | I-V-T measurements |

4. Results and discussion

The '2-omega' experimental data on GaAs and STO at room temperature is summarized in Fig. 2 and Fig. 3 respectively. Fig. 2 also shows the prediction of the deviation from the Fourier law if the accumulation function were as reported by Ref. [2] for GaAs. These predictions were generated by discretizing the reported accumulation function into five channels and applying a multi-band formulation of the EFL [27]. This prediction is more accurate than that of Ref. [27] because fully numerical solutions of the EFL equations were used in this paper, as opposed to an approximate analytical solution. Either prediction falls significantly short of experimental findings.

For GaAs, we extract the following parameters of the MFP accumulation function from the EFL: $\kappa^{LF} = 38 \pm 1$ W/m-K and $\Lambda = 4 \pm 0.5$ micron. From this information, we can reconstruct two points of the accumulation function, as follows [27]: (a) The conductivity of the Fourier law (diffusive) modes $\kappa^{HF} = (\kappa^{bulk} - \kappa^{LF})$ corresponds to mean-free paths of less than 0.37 (or 1/e) times the smallest heater-thermometer separation, which is 1096/e = 403 nm in this experiment. (b) $\Lambda$ corresponds to the point where the conductivity accumulates to 88% of its bulk value [27].

In Fig. 4, we plot the accumulation function of GaAs corresponding to this parameters set and normalized to the bulk conductivity (60 W/m-K) together with the accumulation function reported by [2]. It is seen that they differ widely. Extraction of such wide swathes of the MFPAF from frequency-domain thermoreflectance data as shown in Fig. 4 was earlier criticized by some of the authors on grounds that the raw experimental data could be fit equally well using the compact two-parameter set provided by the EFL (please see Ref. [21]), where the point was also made that the underlying assumption (due to Koh and Cahill [25]) that phonons of MFP greater than the thermal penetration depth are entirely lost to the measurement was *prima facie* incorrect. Despite these criticisms, the agreement of the accumulation function with the first-principles calculations of [26] provides a strong argument in favor of the veracity of the report of [2]. The reason behind the divergence of our results from those of [2] is not known at the time of this writing, and warrants further study.

The thermal conductivity of bulk STO accumulates to its bulk value within ~100 nm [20]. Therefore, our micron-scale experiment should give a null result (no deviation from the Fourier law within experimental error). That this is indeed the case is evident from Fig. 3. The deviations are small relative to the temperature oscillation amplitude (7-10 K), and even negative deviation is seen for certain separations.

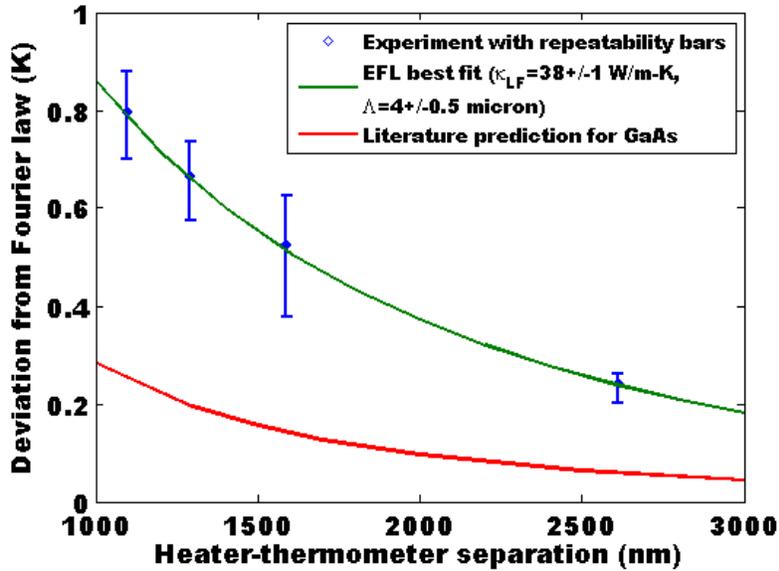

*Fig. 2: Deviation of the experimental temperature oscillation amplitude from the Fourier law prediction for GaAs at 300 K, plotted as a function of heater-thermometer center-to-center separation. The data is fit to the enhanced Fourier law (EFL) and the parameters of the accumulation function are extracted as $\kappa^{LF} = 38 \pm 1$ W/m-K and $\Lambda = 4 \pm 0.5$ micron. The uncertainty in parameter values reflects the repeatability bars shown in the figure. Also shown is the predicted curve using the accumulation function reported in the literature [2][26]*

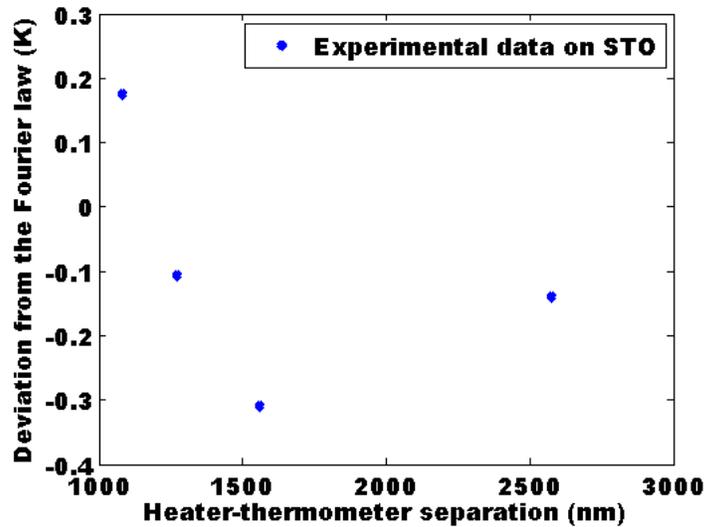

*Fig. 3: Deviation of the experimental temperature oscillation amplitude from the Fourier law prediction for strontium titanate at 300 K, plotted as a function of heater-thermometer center-to-center separation. The data shows no clear trend, and the deviation is small relative to the temperature oscillation amplitude, the latter being on the order of 7-10 K.*

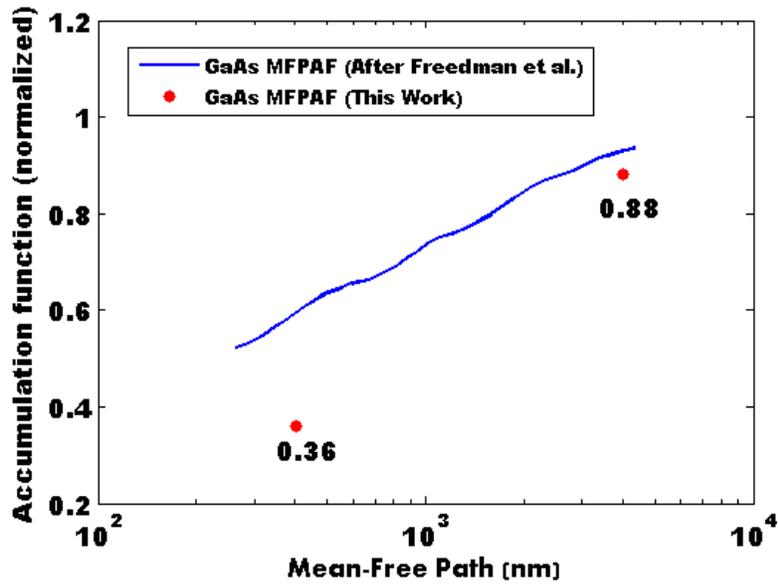

*Fig. 4: Phonon mean-free path accumulation function for GaAs at room temperature. The current work shows that only 36% of the bulk thermal conductivity may be attributed to phonons of MFP less than 400 nm, whereas Freedman et al. [2] estimates this at 60%. After Freedman et al., [2]*

The 50 nm oxide layer (deposited for electrical isolation of the metal lines from the unintentionally conducting substrates) as well as the effects of the finite thickness of the heater line have been ignored in the analysis. The oxide layer may cause a thermal impedance and an accompanying temperature drop between the thermometer and the substrate, reducing the actual temperature of the material surface. The finite thickness of the heater line may cause some heat storage in the heater so that the power fluxed into the material may not equal the power input. Also, the finite thickness of the thermometer line may result in a thermal impedance between it and the material under test.

Fig. 5 shows Fourier law simulations including as well as excluding all three effects, assuming 500 nm wide, 1.2 micron thick metal lines separated by 1000 nm (center-center) deposited on 50 nm SiO2 deposited on GaAs. The thermal conductivity of the oxide is taken to be 0.67 W/m-K as estimated by fitting to 3-omega data on GaAs. This deviates from the nominal oxide thermal conductivity of 1.1 W/m-K in that it includes the boundary thermal impedance between the heater and oxide, and between oxide and GaAs. The thermal conductivity of the metal line is taken as 72 W/m-K, in accord with the accepted value for Pt. Where the heater and thermometer line thickness are excluded, a heat-flux boundary condition is used, which implies that all the input power is fluxed into the material, with no storage in the heater. It is clear from Fig. 5 that all three effects mentioned above are negligible, altogether causing a net error of ~4% in the temperature oscillation amplitude.

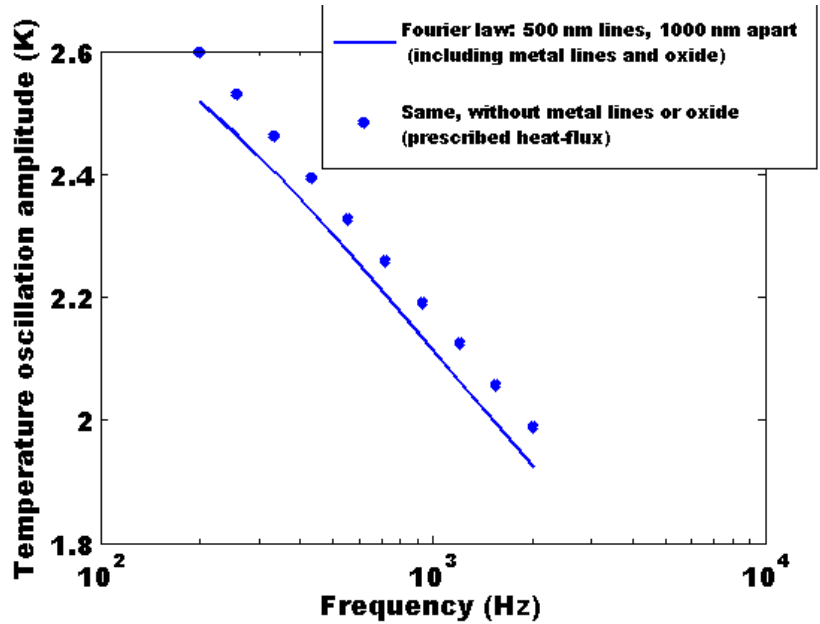

*Fig. 5: Fourier law simulations including and excluding the finite thickness of the heater and thermometer lines, and the effect of the 50 nm SiO$_2$ layer.*

The finite length of the heater and thermometer lines (300 micron) may cause three-dimensional edge effects neglected in our 2-D analysis. In Fig. 6 we plot a sample experimental dataset together with 2D and 3D simulations of the Fourier law as well as the EFL. The metal lines on 50 nm SiO$_2$ on GaAs are 652 nm wide and ~1.2 micron thick, separated center-to-center by 1094 nm. The parameters used for 2D and 3D EFL simulations are $\kappa^{LF} = 38$ W/m-K and $\Lambda = 4$ micron as derived above. Please see Table 1 for the remaining parameters.

We see that 3D effects are negligible at the frequency used throughout in the analysis, 2000 Hz. However, 3D effects are by no means negligible at 200 Hz. (Below 200 Hz, effects due to the finite thickness of the substrate become important, and therefore simulations are not carried out at lower frequencies.). We conjecture the possibility of using frequency-dependent 2-omega data to refine the accumulation function to yield a series of small steps instead of two points (Fig. 4), since it is seen that the 200 Hz data could not be fit adequately by 3D simulations that utilized the EFL parameter set derived at 2000 Hz. However, we were dissuaded from doing so owing to the extreme computational expense incurred by 3D simulations, especially by the iterative optimization routine used to derive best-fit parameters from the experimental data.

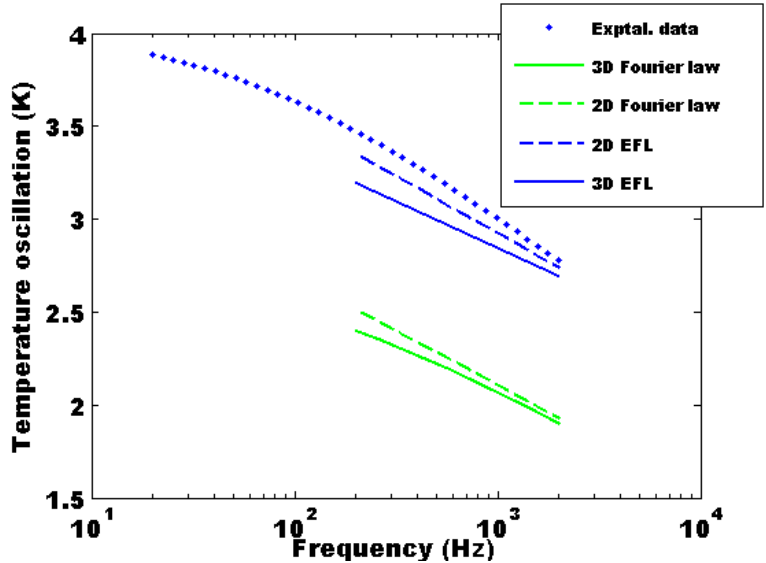

*Fig. 6: Sample experimental data, together with 2D and 3D analyses using both the Fourier law and the enhanced Fourier law (EFL). Please see text for details.*

5. Conclusions

In conclusion, a novel purely electrical probe of the mean-free path accumulation function has been proposed with notable advantages over traditional optical methods, namely insensitivity to interfacial thermal impedances and compatibility with temperature-controlled chambers that lack optical ports. Significant deviations from the Fourier law have been noted for this experiment when conducted on bulk GaAs, and the MFP spectrum of GaAs is seen to be much wider than previously reported. 2-omega experiments analyzed using the enhanced Fourier law suggest that phonons in GaAs with MFP up to 400 nm carry just 36% of the bulk thermal conductivity. This experiment being sensitive to the accumulation function in the region spanning several hundreds of nm to several microns, forms a useful tool along with transient grating and frequency-domain thermoreflectance experiments in mapping out the accumulation function of bulk matter.

Acknowledgments:

We wish to thank Dr. Sean Lubner and Professor Chris Dames (University of California Berkeley) for helpful discussions, Professor Jonathan Malen and Dr. Justin Freedman (Carnegie Mellon University, USA) for kindly supplying us with GaAs MFP spectral data used in Fig. 4, and Jiayi Jiang and Dr. Nathan Weitzner (University of California Santa Barbara) for photolithography mask design and preparation. This work was funded by the National Science Foundation, USA under project number CMMI-1363207.

Appendix

Process for fabricating devices for testing

In this appendix, we describe the device fabrication process in detail in order to facilitate replication of our results.

(a) On GaAs wafers:

50 nm $SiO_2$ was deposited by plasma-enhanced chemical vapor deposition (PECVD) at 250 C on a clean, double-side polished 2-inch diameter unintentionally doped GaAs wafer. The wafer was coated with photoresist NR9-1000 from Futurrex®, and spread to a uniform thickness of 1300 nm by spinning at 2000 rpm. After a 135 C bake for 3 min, it was exposed to UV radiation under a photo-mask for 0.92 sec in a GCA Auto-stepper. A post-develop bake was performed at 100C for 2 min, followed by resist development by exposure to developer MF726 from MicroChemicals® for 20 sec. The development was completed with an $O_2$ plasma exposure for 30 sec to de-scum developed areas. A Ti adhesion layer 23 nm thick was deposited on the developed wafer, followed by a 1200 nm Al layer, both deposited by electron-beam (e-beam) evaporation. Finally, the metal was lifted off from undeveloped areas using 1165 stripper from MicroPosit®.

(b) On strontium titanate (STO) wafers:

An STO wafer 30 mm in diameter was cleaned and 50 nm $SiO_2$ was deposited by radio-frequency sputtering, since the wafer could not withstand the 250 C temperature of the PECVD chamber. The deposition conditions were as follows: RF power = 250W, DC Bias = 150 Volts, $O_2$ flow rate = 2.5 sccm, Ar flow rate = 25 sccm, silicon target as the source of silicon, pressure = 3 mTorr, temperature = 20 C. Since the STO wafer was transparent, the resist deposited directly on $SiO_2$ failed to develop, and the wafer had to be rendered opaque to enable resist deposition. This was accomplished by deposition of 20/10 nm Ti/Au layer on the oxide by e-beam evaporation. The resist deposition and patterning proceeded exactly as for the GaAs wafer. This was followed by e-beam metal deposition (Ti/Au 20/600 nm) and lift-off in 1165 stripper. This metal layer served both as a hard mask for removing the original Ti/Au layer wherever it was not needed, as well as the heater/thermometer metallization. The original Ti/Au 20/10 nm layer was etched away using inductively-coupled plasma (ICP) etch for 90 sec. The ICP etch conditions were as follows: $Cl_2$ flow rate = 44 sccm, Ar flow rate = 20 sccm, pressure = 1.3 Pa, RF power = 200 W, DC Bias = 50 V. Etch rate calibrations showed that the Ti etch rate = 40 nm/min, Au etch rate = 20 nm/min, $SiO_2$ etch rate= 40 nm/min and SrTiO3 etch rate = 4nm/min. Thus the 90 sec ICP etch removed the Ti and Au layers between the heater and thermometer lines with minimal to no invasion of the STO substrate.